\documentclass[10pt,conference]{IEEEtran}

\usepackage[english]{babel} % Different languages:
\usepackage{amsmath}                % AMS Fonts
\usepackage{amsfonts}               % AMS Fonts 
\usepackage{amssymb}                % Mathemat. symbols of AMS
\usepackage{amsopn}                 % Further math. operators like
\allowdisplaybreaks[4] % Page breaks within eqnarray and IEEEeqnarray
\usepackage{bbm}                    % $\mathbbm{1}$ yields a special ``1''
\usepackage{mathrsfs}               % new mathfonts: \mathscr{ABC}
\usepackage{calc}                   % Calculations within the tex-file
\usepackage[dvips]{graphicx}        % Further graphic-commands (use
\usepackage{epsfig}                % Easier syntax for positioning of
\usepackage{psfrag}                 % Include PS-graphics with LaTeX-
\usepackage[dvips]{color}           % Coloring of text
\usepackage{fancyhdr}               % Special page layouts
\usepackage{verbatim}               % citations directly included
\usepackage{exscale}                % Improving on a weakness of
\usepackage{research5}
\newtheorem{theorem}{Theorem}

\title{Multipath Channels of Bounded Capacity}

\author{\authorblockN{Tobias Koch ~~~ Amos Lapidoth}
\authorblockA{ETH
  Zurich, Switzerland\\
Email: \{tkoch, lapidoth\}@isi.ee.ethz.ch}}

\begin{document}

\maketitle

\begin{abstract}
The capacity of discrete-time, non-coherent, multipath fading channels
is considered. It is shown that if the delay spread is large in the
sense that the
variances of the path gains do not decay faster than geometrically,
then capacity is bounded in the signal-to-noise ratio.
\end{abstract}

\section{Introduction}
\label{sec:intro}
This paper studies non-coherent multipath (frequency-selective) fading
channels. Such channels have been investigated extensively in the
wideband regime where the signal-to-noise ratio (SNR) is typically
small, and it was shown that in the limit as the available bandwidth
tends to infinity the capacity of the fading channel is the same as the
capacity of the additive white Gaussian noise (AWGN)
channel of equal received power, see \cite{gallager68}.\footnote{However, in
  contrast to the infinite bandwidth
  capacity of the AWGN channel where the conditions on the capacity
  achieving input distribution are not so stringent, the infinite
  bandwidth capacity of non-coherent fading channels can only be achieved by
  signaling schemes which are ``peaky''; see also
  \cite{sethuramanwanghajeklapidoth07}, \cite{medardgallager02}, \cite{telatartse00} and
  references therein.}

When the SNR is large we encounter a different situation.
Indeed, it has been shown in \cite{lapidothmoser03_3} for non-coherent \emph{frequency-flat} fading channels
that if the fading process is regular in the sense that the present
fading cannot be predicted perfectly from its past, then at high SNR
capacity only increases double-logarithmically in the SNR.  This is in
stark contrast to the logarithmic growth of the AWGN capacity. See \cite{kochlapidoth05_1},
\cite{kochlapidoth05_3}, \cite{lapidothmoser06}, and \cite{moser07} for
extensions to multi-antenna systems, and see \cite{lapidoth05} and
\cite{kochlapidoth06} for extensions to non-regular fading, i.e., when
the present fading can be predicted perfectly from its past. Thus,
communicating over non-coherent flat-fading channels at high SNR is
power inefficient.

In this paper, we show that communicating over non-coherent \emph{multipath}
fading channels at high SNR is not merely power inefficient, but even
worse: if
the delay spread is large in the sense that the variances of the path gains do not decay faster
than geometrically, then capacity is \emph{bounded} in the SNR. For
such channels, capacity does not tend to infinity as the SNR tends to
infinity. To state this result precisely we begin with a mathematical
description of the channel model.

\subsection{Channel Model}
\label{sub:channelmodel}
Let $\Complex$ and $\Integers^+$ denote the set of complex numbers and
the set of positive integers, respectively.
We consider a discrete-time multipath fading channel whose channel
output $Y_k \in \Complex$ at time $k \in \Integers^+$ corresponding to
the channel inputs $(x_1,x_2,\ldots,x_k)\in\Complex^k$ is given by
\begin{equation}
  Y_k = \sum_{\ell=1}^k H_k^{(k-\ell)}x_{\ell}+Z_k. \label{eq:channel}
\end{equation}
Here, $H_k^{(\ell)}$ denotes the time-$k$ gain of the $\ell$-th path,
and $\{Z_k\}$ is a sequence of independent and identically
distributed (IID), zero-mean, variance-$\sigma^2$,
circularly-symmetric, complex Gaussian
random variables. We assume that for each path $\ell\in\Integers_0^+$
(with $\Integers_0^+$ denoting the set of non-negative integers)
 the stochastic process $\big\{H_k^{(\ell)}, k\in\Integers^+\big\}$  is a zero-mean stationary
process. We denote its variance and its differential entropy rate by
\begin{equation*}
  \alpha_{\ell} \triangleq \E{\big|H_k^{(\ell)}\big|^2}, \qquad \ell\in\Integers_0^+
\end{equation*}
and
\begin{equation*}
  h_{\ell} \triangleq
  \lim_{n\to\infty}\frac{1}{n}h\left(H_{1}^{(\ell)},H_2^{(\ell)},\ldots,H_n^{(\ell)}\right),
  \quad \ell \in\Integers_0^+,
\end{equation*}
respectively. We further assume that
\begin{equation}
  \sup_{\ell\in\Integers_0^+} \alpha_{\ell} < \infty \qquad \textnormal{and} \qquad
  \inf_{\ell\in\set{L}} h_{\ell}
  > -\infty,\label{eq:entropyrate}
\end{equation}
where the set $\set{L}$ is defined as $\set{L}\triangleq
\{\ell\in\Integers_0^+:\alpha_{\ell}>0\}$.
We finally assume that the processes
\begin{equation*}
\big\{H_k^{(0)},
k\in\Integers^+\big\}, \big\{H_k^{(1)},
k\in\Integers^+\big\},\ldots
\end{equation*}
are independent (``uncorrelated scattering''), that they are
jointly independent of $\{Z_k\}$, and that the joint law of
\begin{equation*}
\big(\{Z_k\},\big\{H_k^{(0)},
k\in\Integers^+\big\}, \big\{H_k^{(1)},
k\in\Integers^+\big\},\ldots\big)
\end{equation*}
does not depend on the input sequence $\{x_k\}$.
 We consider a \emph{non-coherent} channel model where
neither the transmitter nor the receiver is cognizant of the realization of
$\big\{H_k^{(\ell)}, k\in\Integers^+\big\}$, $\ell\in\Integers_0^+$,
but both are aware of their statistic. We do not assume that the path
gains are Gaussian.

\subsection{Channel Capacity}
\label{sub:capacity}
Let $A_m^n$ denote the sequence $A_m,A_{m+1},\ldots,A_n$. We define the \emph{capacity} as
\begin{equation}
  C(\SNR) \triangleq \varliminf_{n \to \infty} \frac{1}{n} \sup
  I\big(X_1^n;Y_1^n\big), \label{eq:capacity}
\end{equation}
where the maximization is over all joint distributions on
$X_1,X_2,\ldots,X_n$ satisfying the power constraint
\begin{equation}
  \frac{1}{n} \sum_{k=1}^n \E{|X_k|^2} \leq \const{P}, \label{eq:power}
\end{equation}
and where $\SNR$ is defined as
\begin{equation}
  \SNR \triangleq \frac{\const{P}}{\sigma^2}.
 \end{equation}
By Fano's inequality, no rate above $C(\SNR)$ is
achievable.\footnote{See \cite{verduhan94} for conditions that
  guarantee that $C(\SNR)$ is achievable.} (See
\cite{coverthomas91} for a definition of an achievable rate.)

Notice that the above channel \eqref{eq:channel} is generally not
stationary\footnote{By a stationary channel we mean a channel where
  for any stationary input $\{X_k\}$ the pair $\{(X_k,Y_k)\}$ is
  jointly stationary.} since
the number of terms (paths) influencing $Y_k$ depends on $k$. It is
therefore \emph{prima facie} not clear whether the liminf on the RHS of
\eqref{eq:capacity} is a limit.

\subsection{Main Result}
\label{sub:result}
\begin{theorem}
  \label{thm:main}
  Consider the above channel model. Then
  \begin{equation}
    \left(\varliminf_{\ell\to\infty}
    \frac{\alpha_{\ell+1}}{\alpha_{\ell}} > 0\right) \;\; \Longrightarrow \;\;\left(
    \sup_{\SNR>0} C(\SNR) <\infty\right),\label{eq:parti}
  \end{equation}
  where we define, for any $a>0$, $a/0\triangleq\infty$ and
  $0/0\triangleq 0$.
\end{theorem}

For example, when $\{\alpha_{\ell}\}$ is a geometric sequence, i.e.,
$\alpha_{\ell} = \rho^{\ell}$, $\ell\in\Integers_0^+$ for some
$0<\rho<1$, then capacity is bounded.

Theorem~\ref{thm:main} is proved in Section~\ref{sec:proof} where it
is even shown that \eqref{eq:parti} would continue to hold if we
replaced the liminf in \eqref{eq:capacity} by a
limsup. Section~\ref{sec:unbounded} addresses briefly multipath
channels of \emph{unbounded} capacity.

\section{Proof of Theorem~\ref{thm:main}}
\label{sec:proof}
The proof follows along the same lines as the proof of \cite[Thm.~1i)]{kochlapidothsotiriadis07_2}.

We first note that it follows from the left-hand side (LHS) of \eqref{eq:parti} that we can find an
$\ell_0 \in \Integers_0^+$ and a $0<\rho<1$ so that $\alpha_{\ell_0}>0$ and
\begin{equation}
  \frac{\alpha_{\ell+1}}{\alpha_{\ell}} \geq \rho, \qquad \ell =
  \ell_0, \ell_0+1,\ldots.
\end{equation}
We continue with the chain rule for mutual information
\begin{IEEEeqnarray}{lCl}
  \frac{1}{n} I(X_1^n;Y_1^n)
  & = & \frac{1}{n} \sum_{k=1}^{\ell_0}
  I\big(X_1^n;Y_k\big|Y_1^{k-1}\big)\nonumber\\
  & & {} + \frac{1}{n}\sum_{k=\ell_0+1}^n
  I\big(X_1^n;Y_k\big|Y_1^{k-1}\big).\IEEEeqnarraynumspace \label{eq:chainrule}
\end{IEEEeqnarray}
Each term in the first sum on the right-hand side (RHS) of
\eqref{eq:chainrule} is upper bounded by\footnote{Throughout this paper, $\log(\cdot)$
  denotes the natural logarithm function.}
\begin{IEEEeqnarray}{lCl}
  \IEEEeqnarraymulticol{3}{l}{I\big(X_1^n;Y_k\big|Y_1^{k-1}\big)}\nonumber\\
  \quad & \leq & h(Y_k) -
  h\Big(Y_k\Big|Y_1^{k-1},X_1^n,H_k^{(0)},H_k^{(1)},\ldots,H_k^{(k-1)}\Big)\nonumber\\
  & \leq & \log\left(\pi
  e\left(\sigma^2+\sum_{\ell=1}^k \alpha_{k-\ell}\E{|X_{\ell}|^2}\right)\right)
  - \log\big(\pi e\sigma^2\big) \nonumber\\
  & \leq & \log\left(1+\sup_{\ell\in\Integers_0^+} \alpha_{\ell} \cdot n
  \cdot \SNR\right), \label{eq:firstl}
\end{IEEEeqnarray}
where the first inequality follows because conditioning cannot increase
entropy; the second inequality follows from the entropy maximizing
property of Gaussian random variables \cite[Thm.~9.6.5]{coverthomas91}; and the last inequality
follows by upper bounding $\alpha_{\ell}\leq
\sup_{\ell\in\Integers_0^+}\alpha_{\ell}$, $\ell=0,1,\ldots,k-1$
and from the power constraint \eqref{eq:power}.

For $k=\ell_0+1,\ell_0+2,\ldots,n$, we upper bound
$I\big(X_1^n;Y_k\big|Y_1^{k-1}\big)$ using the general upper bound for
mutual information \cite[Thm.~5.1]{lapidothmoser03_3}
\begin{equation}
  I(X;Y) \leq \int D\big(W(\cdot|x)\big\| R(\cdot)\big) \d Q(x), \label{eq:duality}
\end{equation}
where $D(\cdot\|\cdot)$ denotes relative entropy, $W(\cdot|\cdot)$ is
the channel law, $Q(\cdot)$ denotes
the distribution on the channel input $X$, and $R(\cdot)$ is any
distribution on the output alphabet.\footnote{For channels with
  finite input and output alphabets this inequality follows by
  Tops{\o}e's identity \cite{topsoe67}; see also
  \cite[Thm.~3.4]{csiszarkorner81}.} Thus, any choice of output
distribution $R(\cdot)$ yields an upper bound on the mutual
information.

For any given $Y_1^{k-1}=y_1^{k-1}$, we choose the output distribution
  $R(\cdot)$ to be of density
\begin{equation}
  \frac{\sqrt{\beta}}{\pi^2 |y_k|}\frac{1}{1+\beta
  |y_k|^2}, \qquad y_k \in \Complex, \label{eq:cauchy}
\end{equation}
with $\beta=1/(\tilde{\beta}|y_{k-\ell_0}|^2)$ and\footnote{When $y_{k-\ell_0}=0$,
  then the density of the Cauchy distribution \eqref{eq:cauchy} is
  undefined. However, this event is of zero probability and has
  therefore no impact on the mutual information $I\big(X_1^n;Y_k\big|Y_1^{k-1}\big)$.}
\begin{equation}
  \tilde{\beta} = \min\left\{\rho^{\ell_0-1}
  \frac{\alpha_{\ell_0}}{\max_{0\leq\ell'\leq\ell_0} \alpha_{\ell'}},
  \alpha_{\ell_0}, \rho^{\ell_0}\right\}.
\end{equation}
With this choice
\begin{equation}
  0 < \tilde{\beta} < 1 \qquad \textnormal{and} \qquad \tilde{\beta} \alpha_{\ell}
  \leq \alpha_{\ell+\ell_0}, \quad \ell \in \Integers_0^+. \label{eq:beta}
\end{equation}
Using \eqref{eq:cauchy} in \eqref{eq:duality}, and averaging over
$Y_1^{k-1}$, we obtain
\begin{IEEEeqnarray}{lCl}
  \IEEEeqnarraymulticol{3}{l}{I\big(X_1^n;Y_k\big|Y_1^{k-1}\big)}\nonumber\\
  \quad & \leq & \frac{1}{2} \E{\log
    |Y_k|^2} + \frac{1}{2}\E{\log\big(\tilde{\beta}
    |Y_{k-\ell_0}|^2\big)} \nonumber\\
  & & {} +
  \E{\log\big(\tilde{\beta}|Y_{k-\ell_0}|^2+|Y_k|^2\big)} - h\big(Y_k\big|X_1^n,Y_1^{k-1}\big) \nonumber\\
  & & {} - \E{\log
  |Y_{k-\ell_0}|^2} + \log \frac{\pi^2}{\tilde{\beta}}.\label{eq:up1}
\end{IEEEeqnarray}

We bound the terms in \eqref{eq:up1} separately. We begin with
\begin{IEEEeqnarray}{lCl}
  \E{\log |Y_k|^2}
  & = & \E{\Econd{\log|Y_k|^2}{X_1^k}}\nonumber\\
  & \leq & \E{\log\left(\Econd{|Y_k|^2}{X_1^k}\right)}\nonumber\\
  & = &\E{ \log\left(\sigma^2+\sum_{\ell=1}^{k} \alpha_{k-\ell}
  |X_{\ell}|^2\right)}, \label{eq:U1}
\end{IEEEeqnarray}
where the inequality follows from Jensen's inequality.
Likewise, we use Jensen's inequality and \eqref{eq:beta} to upper bound
\begin{IEEEeqnarray}{lCl}
  \E{\log\big(\tilde{\beta}
      |Y_{k-\ell_0}|^2\big)}
  & \leq & 
  \E{\log\left(\!\tilde{\beta}\sigma^2+\tilde{\beta}
  \sum_{\ell=1}^{k-\ell_0}\alpha_{k-\ell_0-\ell} |X_{\ell}|^2\!\right)\!}
  \nonumber\\
  & \leq &
  \E{\log\left(\sigma^2+\sum_{\ell=1}^{k-\ell_0}\alpha_{k-\ell}
  |X_{\ell}|^2\right)} \label{eq:U2}
\end{IEEEeqnarray}
and
\begin{IEEEeqnarray}{lCl}
  \IEEEeqnarraymulticol{3}{l}{\E{\log\big(\tilde{\beta}|Y_{k-\ell_0}|^2+|Y_k|^2\big)}}\nonumber\\
  \;\; & \leq & \E{\log\left(2\sigma^2+2\sum_{\ell=1}^{k-\ell_0} \alpha_{k-\ell}|X_{\ell}|^2+\!\!\sum_{\ell=k-\ell_0+1}^{k}\!\! \alpha_{k-\ell}|X_{\ell}|^2\right)}\nonumber\\
  & \leq & \log 2 + \E{\log\left(\sigma^2+\sum_{\ell=1}^k \alpha_{k-\ell}|X_{\ell}|^2\right)},\label{eq:U3}
\end{IEEEeqnarray}
where the second inequality follows because
$\sum_{\ell=k-\ell_0+1}^{k}\alpha_{k-\ell}|X_{\ell}|^2 \leq 2 \sum_{\ell=k-\ell_0+1}^{k}\alpha_{k-\ell}|X_{\ell}|^2$.

Next, we derive a lower bound on
$h\big(Y_k\big|X_1^n,Y_1^{k-1}\big)$. Let $\vect{H}_{k'} =
\big(H_{k'}^{(0)},H_{k'}^{(1)},
\ldots,H_{k'}^{(k-1)}\big)$, $k'=1,2,\ldots,k-1$. We have
\begin{IEEEeqnarray}{lCl}
  h\big(Y_k\big|X_1^n,Y_1^{k-1}\big) & \geq &
  h\big(Y_k\big|X_1^n,Y_1^{k-1},\vect{H}_{1}^{k-1}\big)\nonumber\\
  & = & h\big(Y_k\Big|X_1^n,\vect{H}_{1}^{k-1}\big),
\end{IEEEeqnarray}
where the inequality follows because conditioning cannot increase
entropy, and where the equality follows because, conditional on
$\big(X_1^n,\vect{H}_{1}^{k-1}\big)$, $Y_k$ is independent of
$Y_1^{k-1}$.
Let $\set{S}_k$ be defined as
\begin{equation}
\set{S}_k\triangleq\{\ell = 1,2,\ldots,k:
\min\{|x_{\ell}|^2,\alpha_{k-\ell}\}>0\}.
\end{equation}
Using the entropy power inequality \cite[Thm.~16.6.3]{coverthomas91},
and using that the processes
\begin{equation*}
\big\{H_{k}^{(0)}, k\in\Integers^+\big\},\big\{H_{k}^{(1)}, k\in\Integers^+\big\},\ldots
\end{equation*}
are independent and jointly independent of $X_1^n$, it can be shown that for any given $X_1^n=x_1^n$
\begin{IEEEeqnarray}{lCl}
  \IEEEeqnarraymulticol{3}{l}{h\left(\left.\sum_{\ell=1}^k
        H_k^{(k-\ell)}X_{\ell}+Z_k\right|X_1^n=x_1^n,\vect{H}_1^{k-1}\right)}\nonumber\\
   & \geq & \log\left(\sum_{\ell\in\set{S}_k}\!
      e^{h\big(H_k^{(k-\ell)}X_{\ell}\big|X_{\ell}=x_{\ell},\{H_{k'}^{(k-\ell)}\}_{k'=1}^{k-1}\big)}\!+e^{h(Z_k)}
    \!\right)\!.\IEEEeqnarraynumspace\label{eq:entropy1}
\end{IEEEeqnarray}
We lower bound the differential entropies on the RHS of
\eqref{eq:entropy1} as follows. The differential entropies in the sum are lower
bounded by
\begin{IEEEeqnarray}{lCl}
  \IEEEeqnarraymulticol{3}{l}{h\left(\left.H_k^{(k-\ell)}X_{\ell}\right|X_{\ell}=x_{\ell},\big\{H_{k'}^{(k-\ell)}\big\}_{k'=1}^{k-1}\right)}\nonumber\\
  \; & = &
  \log\big(\alpha_{k-\ell}|x_{\ell}|^2\big)+h\!\left(\!\left.H_k^{(k-\ell)}\right|\!\big\{H_{k'}^{(k-\ell)}\big\}_{k'=1}^{k-1}\right)\!-\log\alpha_{k-\ell}\nonumber\\
  & \geq &
  \log\big(\alpha_{k-\ell}|x_{\ell}|^2\big) +
  \inf_{\ell\in\set{L}}\left(h_{\ell}-\log\alpha_{\ell}\right), \qquad \ell\in\set{S}_k,\label{eq:entropy2}
\end{IEEEeqnarray}
where the equality follows from the behavior of differential entropy under
scaling; and where the inequality follows by the stationarity of the process
$\big\{H_k^{(k-\ell)},k\in\Integers^+\big\}$
which implies that the differential entropy
$h\big(H_k^{(k-\ell)}\big|\big\{H_{k'}^{(k-\ell)}\big\}_{k'=1}^{k-1}\big)$
cannot be smaller than the differential entropy rate $h_{k-\ell}$
\cite[Thms.~4.2.1 \& 4.2.2]{coverthomas91}, and by lower bounding
$(h_{k-\ell}-\log\alpha_{k-\ell})$ by
$\inf_{\ell\in\set{L}}(h_{\ell}-\log\alpha_{\ell})$ (which holds for each
$\ell\in\set{S}_k$ because $\set{S}_k\subseteq\set{L}$).
The last differential entropy on the RHS of \eqref{eq:entropy1} is
lower bounded by
\begin{equation}
  \label{eq:obvious}
  h(Z_k) \geq \inf_{\ell\in\set{L}}\left(h_{\ell}-\log\alpha_{\ell}\right)
  + \log\sigma^2,
\end{equation}
which follows by noting that
\begin{equation}
  \inf_{\ell\in\set{L}}\left(h_{\ell}-\log\alpha_{\ell}\right) \leq \log(\pi e).
\end{equation}
Applying \eqref{eq:entropy2} \& \eqref{eq:obvious} to
\eqref{eq:entropy1}, and averaging over $X_1^n$, yields then
\begin{IEEEeqnarray}{lCl}
  h\big(Y_k\big|X_1^n,Y_1^{k-1}\big)
  & \geq & \E{\log\left(\sigma^2+\sum_{\ell=1}^k \alpha_{k-\ell}
      |X_{\ell}|^2\right)} \nonumber\\
  & & {}+ \inf_{\ell\in\set{L}}\left(h_{\ell}-\log\alpha_{\ell}\right).\IEEEeqnarraynumspace
  \label{eq:U4}
\end{IEEEeqnarray}

We continue with the analysis of \eqref{eq:up1} by lower bounding
$\E{\log|Y_{k-\ell_0}|^2}$. To this end, we write the
expectation as
\begin{equation*}
  \E{\Econd{\log\left|\sum_{\ell=1}^{k-\ell_0}H_{k-\ell_0}^{(k-\ell_0-\ell)}X_{\ell}+Z_{k-\ell_0}\right|^2}{X_1^{k-\ell_0}}}
\end{equation*}
and lower bound the conditional expectation by
\begin{IEEEeqnarray}{lCl}
  \IEEEeqnarraymulticol{3}{l}{\Econd{\log\left|\sum_{\ell=1}^{k-\ell_0}H_{k-\ell_0}^{(k-\ell_0-\ell)}X_{\ell}+Z_{k-\ell_0}\right|^2}{X_1^{k-\ell_0}=x_1^{k-\ell_0}}} \nonumber\\
    \quad & = &  \log\left(\sigma^2+\sum_{\ell=1}^{k-\ell_0}\alpha_{k-\ell_0-\ell}|x_{\ell}|^2\right)\nonumber\\
    & & 
    {} - 2\cdot
    \E{\log\left|\frac{\sum_{\ell=1}^{k-\ell_0}H_{k-\ell_0}^{(k-\ell_0-\ell)}x_{\ell}+Z_{k-\ell_0}}{\sqrt{\sigma^2+\sum_{\ell=1}^{k-\ell_0}\alpha_{k-\ell_0-\ell}|x_{\ell}|^2}}\right|^{-1}}
    \nonumber\\
    & \geq &
    \log\left(\sigma^2+\sum_{\ell=1}^{k-\ell_0}\alpha_{k-\ell_0-\ell}|x_{\ell}|^2\right)
    +\log\delta^2 -2 \eps(\delta,\eta)\nonumber\\
    & & {}- \frac{2}{\eta}
    h^{-}\left(\frac{\sum_{\ell=1}^{k-\ell_0}
    H_{k-\ell_0}^{(k-\ell_0-\ell)}x_{\ell}+Z_{k-\ell_0}}{\sqrt{\sigma^2+\sum_{\ell=1}^{k-\ell_0}\alpha_{k-\ell_0-\ell}|x_{\ell}|^2}}\right) \label{eq:bla1}
\end{IEEEeqnarray}
for some $0< \delta\leq 1$ and $0<\eta<1$, where
\begin{equation}
  h^{-}(X) \triangleq \int_{\{x\in\Complex:f_X(x)>1\}} f_X(x)\log f_X(x)\d x,
\end{equation}
and where
$\eps(\delta,\eta)>0$ tends to zero as $\delta \downarrow 0$. (We
write $x_{\ell}$ in lower case to indicate that expectation and entropy are
conditional on $X_1^{k-\ell_0}=x_1^{k-\ell_0}$.) Here,
the inequality follows by writing the expectation in the form
$\E{\log|A|^{-1}\cdot I\{|A|>\delta\}}+\E{\log|A|^{-1}\cdot I\{|A|\leq\delta\}}$
(where $I\{\cdot\}$ denotes the indicator function), and by upper bounding then
the first expectation by $-\log\delta$ and the second expectation using
\cite[Lemma 6.7]{lapidothmoser03_3}. We
continue by upper bounding
\begin{IEEEeqnarray}{lCl}
  \IEEEeqnarraymulticol{3}{l}{h^{-}\left(\frac{\sum_{\ell=1}^{k-\ell_0}
          H_{k-\ell_0}^{(k-\ell_0-\ell)}x_{\ell}+Z_{k-\ell_0}}{\sqrt{\sigma^2+\sum_{\ell=1}^{k-\ell_0}\alpha_{k-\ell_0-\ell}|x_{\ell}|^2}}\right)}\nonumber\\
  \quad & = & h^{+}\left(\frac{\sum_{\ell=1}^{k-\ell_0}
        H_{k-\ell_0}^{(k-\ell_0-\ell)}x_{\ell}+Z_{k-\ell_0}}{\sqrt{\sigma^2+\sum_{\ell=1}^{k-\ell_0}\alpha_{k-\ell_0-\ell}|x_{\ell}|^2}}\right)\nonumber\\
  & & {} - h\left(\frac{\sum_{\ell=1}^{k-\ell_0}
        H_{k-\ell_0}^{(k-\ell_0-\ell)}x_{\ell}+Z_{k-\ell_0}}{\sqrt{\sigma^2+\sum_{\ell=1}^{k-\ell_0}\alpha_{k-\ell_0-\ell}|x_{\ell}|^2}}\right)\nonumber\\
  & \leq & \frac{2}{e}+\log(\pi e) +
  \log\left(\sigma^2+\sum_{\ell=1}^{k-\ell_0}
    \alpha_{k-\ell_0-\ell}|x_{\ell}|^2\right) \nonumber\\
  & & {} - h\left(\sum_{\ell=1}^{k-\ell_0}
      H_{k-\ell_0}^{(k-\ell_0-\ell)}x_{\ell}+Z_{k-\ell_0}\right), \label{eq:blablabla}
\end{IEEEeqnarray}
where $h^{+}(X)$ is defined as $h^+(X)\triangleq h(X)+h^-(X)$.
Here, we applied \cite[Lemma 6.4]{lapidothmoser03_3} to upper bound
\begin{equation}
  h^{+}\left(\frac{\sum_{\ell=1}^{k-\ell_0}
        H_{k-\ell_0}^{(k-\ell_0-\ell)}x_{\ell}+Z_{k-\ell_0}}{\sqrt{\sigma^2+\sum_{\ell=1}^{k-\ell_0}\alpha_{k-\ell_0-\ell}|x_{\ell}|^2}}\right)
        \leq \frac{2}{e}+\log(\pi e).
\end{equation}
Averaging \eqref{eq:blablabla} over
$X_1^{k-\ell_0}$ yields
\begin{IEEEeqnarray}{lCl}
  \IEEEeqnarraymulticol{3}{l}{h^{-}\left(\left.\frac{\sum_{\ell=1}^{k-\ell_0}
          H_{k-\ell_0}^{(k-\ell_0-\ell)}X_{\ell}+Z_{k-\ell_0}}{\sqrt{\sigma^2+\sum_{\ell=1}^{k-\ell_0}\alpha_{k-\ell_0-\ell}|X_{\ell}|^2}}\right|X_1^{k-\ell_0}\right)}\nonumber\\
  \quad & \leq & \frac{2}{e}+\log(\pi e) +
  \E{\log\left(\sigma^2+\sum_{\ell=1}^{k-\ell_0}
      \alpha_{k-\ell_0-\ell}|X_{\ell}|^2\right)} \;\;\nonumber\\
  & & {} - h\left(\left.\sum_{\ell=1}^{k-\ell_0}
      H_{k-\ell_0}^{(k-\ell_0-\ell)}X_{\ell}+Z_{k-\ell_0}\right|X_1^{k-\ell_0}\right)\nonumber\\
  & \leq & \frac{2}{e}+\log(\pi e) - \inf_{\ell\in\set{L}}\left(h_{\ell}-\log\alpha_{\ell}\right), \label{eq:bla2}
\end{IEEEeqnarray}
where the second inequality follows by conditioning the differential entropy
additionally on
$Y_1^{k-\ell_0-1}$, and by using then lower bound \eqref{eq:U4}. 
A lower bound on $\E{\log |Y_{k-\ell_0}|^2}$ follows now by averaging
\eqref{eq:bla1} over $X_1^{k-\ell_0}$, and by applying \eqref{eq:bla2}
\begin{IEEEeqnarray}{lCl}
  \E{\log |Y_{k-\ell_0}|^2}
    & \geq &
    \E{\log\left(\sigma^2+\sum_{\ell=1}^{k-\ell_0}\alpha_{k-\ell_0-\ell}|X_{\ell}|^2\right)}\nonumber\\
    & & {} +\log\delta^2-2\eps(\delta,\eta)-\frac{2}{\eta}\left(\frac{2}{e}+\log(\pi
      e)\!\right) \nonumber\\
    & & {}+ \inf_{\ell\in\set{L}}\frac{2}{\eta} \left(h_{\ell}-\log\alpha_{\ell}\right).\label{eq:U5}
\end{IEEEeqnarray}

Returning to the analysis of \eqref{eq:up1}, we obtain from
\eqref{eq:U5}, \eqref{eq:U4}, \eqref{eq:U3}, \eqref{eq:U2}, and
\eqref{eq:U1}
\begin{IEEEeqnarray}{lCl}
  \IEEEeqnarraymulticol{3}{l}{I\big(X_1^n;Y_k\big|Y_1^{k-1}\big)}\nonumber\\
  \;\; & \leq & \frac{1}{2}\E{\log\left(\sigma^2+\sum_{\ell=1}^k
      \alpha_{k-\ell}|X_{\ell}|^2\right)}\nonumber\\
  & & {} +\frac{1}{2}\E{\log\left(\sigma^2+\sum_{\ell=1}^{k-\ell_0}\alpha_{k-\ell}|X_{\ell}|^2\right)}
  \nonumber\\
  & & {} +\log 2 + \E{\log\left(\sigma^2+\sum_{\ell=1}^k
      \alpha_{k-\ell}|X_{\ell}|^2\right)} \nonumber\\
  & & {} - \E{\log\left(\sigma^2+\sum_{\ell=1}^{k}
      \alpha_{k-\ell}|X_{\ell}|^2\right)} - \inf_{\ell\in\set{L}}\left(h_{\ell}-\log\alpha_{\ell}\right)\nonumber\\
  & & {} -\E{\log\left(\sigma^2+\sum_{\ell=1}^{k-\ell_0}
      \alpha_{k-\ell_0-\ell}|X_{\ell}|^2\right)} \nonumber\\
  & & {} - \log\delta^2 +
  2\eps(\delta,\eta)+\frac{2}{\eta}\left(\frac{2}{e}+\log(\pi
    e)\!\right)\nonumber\\
  & & {} - \inf_{\ell\in\set{L}}\frac{2}{\eta}\left(h_{\ell}-\log\alpha_{\ell}\right)
  +\log\frac{\pi^2}{\tilde{\beta}}\nonumber\\
  & \leq & \const{K} +\E{\log\left(\sigma^2+\sum_{\ell=1}^{k}
      \alpha_{k-\ell} |X_{\ell}|^2\right)}\nonumber\\
  & & {} -
  \E{\log\left(\sigma^2+\sum_{\ell=1}^{k-\ell_0}\alpha_{k-\ell_0-\ell}
  |X_{\ell}|^2\right)}, \label{eq:almostfinished}
\end{IEEEeqnarray}
with
\begin{IEEEeqnarray}{lCl}
  \const{K} & \triangleq & -
  \inf_{\ell\in\set{L}}\left(1+\frac{2}{\eta}\right)
  \left(h_{\ell}-\log\alpha_{\ell}\right)+
  \log\frac{2\pi^2}{\tilde{\beta}\delta^2}\nonumber\\
  & & {} + 2\eps(\delta,\eta) +
  \frac{2}{\eta}\left(\frac{2}{e}+\log(\pi e)\!\right).
\end{IEEEeqnarray}
The second inequality in \eqref{eq:almostfinished} follows because
$\sum_{\ell=1}^{k-\ell_0} \alpha_{k-\ell}|X_{\ell}|^2\leq
\sum_{\ell=1}^k \alpha_{k-\ell}|X_{\ell}|^2$.

In order to show that the capacity is bounded in the $\SNR$, we apply
\eqref{eq:almostfinished} and \eqref{eq:firstl} to
\eqref{eq:chainrule} and use then that for any sequences $\{a_k\}$ and $\{b_k\}$
\begin{IEEEeqnarray}{lCl}
  \sum_{k=\ell_0+1}^n (a_k-b_k) & = & \sum_{k=n-\ell_0+1}^n
 (a_k-b_{k-n+2\ell_0})\nonumber\\
 & & {} + \sum_{k=\ell_0+1}^{n-\ell_0}(a_k-b_{k+\ell_0}).\IEEEeqnarraynumspace \label{eq:sum2}
\end{IEEEeqnarray}
Defining
\begin{equation}
  a_k \triangleq \E{\log\left(\sigma^2+\sum_{\ell=1}^{k}
  \alpha_{k-\ell} |X_{\ell}|^2\right)}
\end{equation}
and
\begin{equation}
  b_k \triangleq \E{\log\left(\sigma^2+\sum_{\ell=1}^{k-\ell_0}\alpha_{k-\ell_0-\ell}
      |X_{\ell}|^2\right)}
\end{equation}
we have for the first sum on the RHS of \eqref{eq:sum2}
\begin{IEEEeqnarray}{lCl}
  \IEEEeqnarraymulticol{3}{l}{\sum_{k=n-\ell_0+1}^n
  (a_k-b_{k-n+2\ell_0})}\nonumber\\
  \quad\; & = & \sum_{k=n-\ell_0+1}^n \E{\log\left(\frac{\sigma^2+\sum_{\ell=1}^k
        \alpha_{k-\ell}
        |X_{\ell}|^2}{\sigma^2+\sum_{\ell=1}^{k-n+\ell_0}\alpha_{k-n+\ell_0-\ell}|X_{\ell}|^2}\right)}\nonumber\\
  & \leq & \ell_0 \log\left(1+\sup_{\ell\in\Integers_0^+}\alpha_{\ell} \cdot n \cdot \SNR\right),\label{eq:last1}
\end{IEEEeqnarray}
which follows by lower bounding the denominator by $\sigma^2$, and by
using then Jensen's inequality along with the last inequality
in \eqref{eq:firstl}. For the
second sum on the RHS of \eqref{eq:sum2} we have
%% \begin{equation}
%%   \sum_{k=\ell_0+1}^{n-\ell_0}(a_k-b_{k+\ell_0}) = 0. \label{eq:last2}
%% \end{equation}
\begin{IEEEeqnarray}{lCl}
  \IEEEeqnarraymulticol{3}{l}{\sum_{k=\ell_0+1}^{n-\ell_0}(a_k-b_{k+\ell_0})}\nonumber\\
  \quad & = & \sum_{k=\ell_0+1}^{n-\ell_0} \E{\log\left(\frac{\sigma^2+\sum_{\ell=1}^k\alpha_{k-\ell}|X_{\ell}|^2}{\sigma^2+\sum_{\ell=1}^k \alpha_{k-\ell}
        |X_{\ell}|^2}\right)}\nonumber\\
 & = & 0.\label{eq:last2}
\end{IEEEeqnarray}
Thus, applying \eqref{eq:almostfinished}--\eqref{eq:last2} and
\eqref{eq:firstl} to \eqref{eq:chainrule}, we obtain
\begin{IEEEeqnarray}{lCl}
  \IEEEeqnarraymulticol{3}{l}{\frac{1}{n}I(X_1^n;Y_1^n)}\nonumber\\
  \;\; & \leq & \frac{\ell_0}{n}
  \log\left(\sigma^2+\sup_{\ell\in\Integers_0^+}\alpha_{\ell} \cdot n \cdot
    \SNR\right) \nonumber\\
  & & {} + \frac{\ell_0}{n}
  \log\left(\sigma^2+\sup_{\ell\in\Integers_0^+}\alpha_{\ell} \cdot n \cdot
    \SNR\right) + \frac{n-2 \ell_0}{n} \const{K},\IEEEeqnarraynumspace
\end{IEEEeqnarray}
which tends to $\const{K}<\infty$ as $n$ tends to infinity. This
proves Theorem~\ref{thm:main}.

\section{Multipath Channels of Unbounded Capacity}
\label{sec:unbounded}
We have seen in Theorem~\ref{thm:main} that if the variances of the path gains $\{\alpha_{\ell}\}$ do not
decay faster than geometrically, then capacity is bounded in the
SNR. In this section, we demonstrate that this need not be the case
when the variances of the path gains
decay faster than geometrically. The following
theorem presents a sufficient condition for the capacity $C(\SNR)$ to
be unbounded in the SNR.

\begin{theorem}
\label{thm:second}
Consider the above channel model. Then
\begin{IEEEeqnarray}{c}
  \label{eq:partii}
  \left(\lim_{\ell\to\infty}\frac{1}{\ell}\log\log\frac{1}{\alpha_{\ell}} = \infty\!\right)
  \Longrightarrow \left(\sup_{\SNR>0} C(\SNR) =\infty\!\right)\!.\IEEEeqnarraynumspace
\end{IEEEeqnarray}
\end{theorem}
\begin{proof}
Omitted.
\end{proof}

\emph{Note:} We do not claim that $C(\SNR)$ is achievable. However, it
can be shown that when, for example, the processes $\big\{H_k^{(\ell)},
k\in\Integers^+\big\}$, $\ell\in\Integers^+_0$
are IID Gaussian, then the maximum achievable rate is unbounded in the
SNR, i.e., any rate is achievable for sufficiently large SNR.

Certainly, the condition on the LHS of \eqref{eq:partii} is satisfied when
the channel has finite memory in the sense that for some finite
$L\in\Integers^+_0$
\begin{equation*}
  \alpha_{\ell} = 0, \qquad \ell=L+1,L+2,\ldots.
\end{equation*}
In this case, \eqref{eq:channel} becomes
\begin{equation}
  Y_k = \left\{\!
  \begin{array}{ll}
    \displaystyle \sum_{\ell=0}^{k-1}
    H_k^{(\ell)}x_{k-\ell}+Z_k, \quad & k=1,2,\ldots,L\\
    \displaystyle \sum_{\ell=0}^{L}
    H_k^{(\ell)}x_{k-\ell}+Z_k, & k=L+1,L+2,\ldots.
  \end{array}\right.\!\!\!\!\!\!\!\label{eq:channel2}
\end{equation}
\newpage
This channel \eqref{eq:channel2} was studied for general (but finite)
$L$ in \cite{kochlapidoth08_2_sub} where it was shown that its
capacity satisfies
\begin{equation}
  \label{eq:preloglog}
  \lim_{\SNR\to\infty} \frac{C(\SNR)}{\log\log\SNR} = 1.
\end{equation}
Thus, for finite $L$, the capacity pre-loglog \eqref{eq:preloglog} is
not affected by the multipath behavior. This is perhaps surprising as
Theorem~\ref{thm:main} implies that if $L=\infty$, and if the
variances of the path
gains do not decay faster than geometrically, then
the pre-loglog is zero.

\section*{Acknowledgment}
Discussions with Helmut B\"olcskei and Giuseppe Durisi are gratefully acknowledged.

%\bibliographystyle{IEEEtran}           % in order of first citation
%\bibliography{/Volumes/Data/tkoch/Library/texmf/tex/bibtex/header_short,/Volumes/Data/tkoch/Library/texmf/tex/bibtex/bibliofile}

\end{document}